\documentclass[12pt]{iopart}
\usepackage[dvipdfmx]{graphicx}
\usepackage{bm}
\usepackage{braket}
\usepackage{iopams} 
\usepackage{cite}
\usepackage{color}
\begin{document}

\title{Coherent two-photon emission from hydrogen molecules excited by counter-propagating laser pulses}

\author{Takahiro~Hiraki$^1$, Hideaki~Hara$^1$, Yuki~Miyamoto$^1$, Kei~Imamura$^1$, Takahiko~Masuda$^1$, Noboru~Sasao$^1$, Satoshi~Uetake$^{1,2}$, Akihiro~Yoshimi$^1$, Koji~Yoshimura$^1$, and Motohiko~Yoshimura$^1$}

\address{$^1$ Research Institute for Interdisciplinary Science, Okayama University, Okayama 700-8530, Japan}
\address{$^2$ PRESTO, Japan Science and Technology, Okayama 700-8530, Japan}
\ead{thiraki@okayama-u.ac.jp}
\vspace{10pt}
\begin{indented}
\item[]May 2018
\end{indented}

\begin{abstract}
We observed two-photon emission signal from the first \textcolor{black}{vibrationally} excited state of parahydrogen gas coherently excited by counter-propagating laser pulses.
A single narrow-linewidth laser source has roles in the excitation of the parahydrogen molecules and the induction of the two-photon emission process. 
We measured  dependences of the signal energy on the detuning, target gas pressure, and input pulse energies.
These results are qualitatively consistent with those obtained by numerical simulations based on Maxwell-Bloch equations with one spatial dimension and one temporal dimension. 
This study of the two-photon emission process in the counter-propagating injection scheme is an important step toward neutrino mass spectroscopy. 
\end{abstract}

%
\vspace{2pc}
%
%
%
%

\section{Introduction}
Emission processes of atoms or molecules could be modified when coherent phenomena \textcolor{black}{are involved in} them. 
A famous example is superradiance, which was first predicted by Dicke~\cite{bib:superradiance}, and has been  observed in various systems~\cite{bib:SR1,bib:SR2,bib:SR3,bib:SR4,bib:SR5,bib:SR6,bib:SR7,bib:SR8}.
In superradiant emission processes, an ensemble of atoms or molecules behaves cooperatively.
Consequently, the emission rate of the ensemble is significantly enhanced compared \textcolor{black}{with} usual cases where each atom or molecule behaves independently~\cite{bib:SRreview}. 
By taking advantage of this enhancement property of superradiance, very weak processes of atomic transitions could be observed~\cite{bib:SRamplification}.

One of the authors of the current paper proposed another coherent amplification scheme~\cite{bib:SPAN2008}.
The principle of this scheme is reviewed in~\cite{bib:ptepreview} and is \textcolor{black}{briefly} described here. 
Let us consider a system where atoms or molecules are excited by lasers to a metastable excited state and then they de-excite with emitting particles. 
In the excitation process, coherence among laser photons is imprinted into the ensemble. 
We denote $\bm{k}_{\rm in}$ $(\bm{k}_{\rm out})$ by the sum of the wave \textcolor{black}{vectors} of excitation (de-excitation) particles. 
The emission rate $R$ of this process is proportional to
\begin{equation}
\left|\int d\bm{r}\sum_{a=1}^{N} \exp({\rm i}(\bm{k}_{\rm in}-\bm{k}_{\rm out})\cdot(\bm{r}-\bm{r}_a))\mathcal{M}(\bm{r}_a)\right|^2, 
\end{equation}
where $a$ denotes an atom or a molecule, $\bm{r}_a$ denotes \textcolor{black}{its} spatial position, and $\mathcal{M}$ represents the transition amplitude of the \textcolor{black}{atom at the position $\bm{r}_a$}. 
If each particle behaves independently, the emission rate is proportional to $N$, which indicates the total number of atoms or molecules involved. 
However, in the case that $\bm{k}_{\rm in}=\bm{k}_{\rm out}$ holds and decoherence time is sufficiently long, the emission rate becomes 
\begin{equation}
R\propto\left|\sum_{a=1}^{N} \mathcal{M}(\bm{r}_a)\right|^2\propto N^2
\end{equation} 
and thus huge rate amplification can be achieved. 
Here rate amplification is realized only when momentum \textcolor{black}{$\bm{k}_{\rm in}=\bm{k}_{\rm out}$, or} conservation among excitation and de-excitation particles holds.

This emission rate amplification method can be applied to resolve fundamental questions in a different context, neutrino physics~\cite{bib:SPAN1,bib:ptepreview}.
In recent neutrino physics, the squared-mass differences and the mixing angles of neutrinos have been measured in various neutrino oscillation experiments~\cite{bib:SKsolar,bib:KamLAND,bib:DayaBay,bib:T2K17,bib:Nova,bib:nuosc}. 
However, the absolute scale of neutrino masses, mass ordering, CP phases, and mass type (Dirac or Majorana) still remain \textcolor{black}{to be explored experimentally}~\cite{bib:directmass,bib:Planck,bib:nufit,bib:DBD}. 
Uncovering these properties is important in terms of particle physics beyond the standard model and cosmology~\cite{bib:nucosmo}. 
We use atomic or molecular de-excitation processes emitting a single photon and a neutrino pair, which we call RENP (Radiative Emission of Neutrino Pair).
In the RENP process, emission rate spectra of the photon have abundant information about these neutrino properties~\cite{bib:observable,bib:boosted}. 
A serious issue when we use RENP processes in experiments is their extremely small rates~\cite{bib:SPAN1,bib:observable,bib:calculation}. 
Thus to amplify the rate by using this amplification method is a key to the observation of RENP processes. 

As of this moment, it is quite difficult to observe RENP processes. One of the reasons of this is our understandings of the rate amplification mechanism is insufficient. 
Although multi-photon emission processes has been extensively studied, most of them did not focus on rate amplification of the processes. Moreover, since multi-photon emission processes are possible dominant background sources in RENP observation experiments, it is necessary to understand in detail the rate amplification in multi-photon emission processes. Thus, we have been studying the coherent amplification mechanism using two-photon emission (TPE) processes. TPE processes are the simplest multi-photon emission processes and observation of TPE processes is easier than those of the RENP or higher-order multi-photon emission processes.

In the previous experiments, we observed the TPE signal from vibrational states of parahydrogen (p-H$_2$) molecules~\cite{bib:TPE1,bib:TPE2,bib:TPE3,bib:TPE4}, which have long decoherence time \textcolor{black}{($\mathcal{O}$(1) ns)}.   
In these experiments two lasers of different colors were injected coaxially in the same direction for the Raman excitation of p-H$_2$. 
In \textcolor{black}{the present} study, in contrast, p-H$_2$ are excited by monochromatic counter-propagating lasers\textcolor{black}{. With this excitation scheme,} 
 another requirement for the amplification of the RENP processes can be satisfied as described below.

We focus on invariant masses of excitation and de-excitation particles in the RENP processes. 
The invariant mass should be conserved in order to amplify the emission rate.
This is because in addition to the momentum conservation condition, energy conservation among excitation and de-excitation particles should also hold. 
The RENP processes occur only when the invariant mass is larger than the sum of the masses of the emitted neutrinos. 

Invariant mass \textcolor{black}{squared} of the two free photons for the excitation is given by 
\begin{equation}
s=2c^2|\bm{p}_1|^2|\bm{p}_2|^2(1-\cos\theta),  
\end{equation} 
where $|\bm{p}_1|$ and $|\bm{p}_2|$ indicate momentum of each photon and $\theta$ indicates the crossing angle between the photons. 
In the case of one-side laser injection scheme ($\theta=0$), $s$ is equal to 0 and the amplification conditions of the RENP process cannot be satisfied.  
In the counter-propagating laser excitation scheme ($\theta=\pi$), in contrast,  
it is possible to satisfy the amplification conditions of the RENP process.
Another merit of the counter-propagating excitation scheme is related to soliton formulation, which is  described in~\cite{bib:soliton, bib:dynamics}. 
For these reasons, to observe and understand the TPE process in the counter-propagating injection scheme is an important step toward observation of RENP processes. 

As in the case of the previous experiments~\cite{bib:TPE1,bib:TPE2,bib:TPE3}, it is helpful for understandings of properties of the coherent amplification to compare actual data to results of the numerical simulation. 
These simulations are based on Maxwell-Bloch equations, which represent developments of the laser fields and the coherence among the ensemble. 
In this paper, we investigate properties of the TPE process in detail through comparisons with simulation results. 
 
The TPE process observed in our previous experiments is a four-wave mixing (FWM) process, which has been studied in nonlinear optics and quantum electronics~\cite{bib:FWMreview}. In the current experiment setup, frequencies of all the four waves are identical. This process is called degenerate four-wave mixing process (DFWM).  
DFWM generation experiments in the counter-propagating scheme~\cite{bib:boyd_DFWM1,bib:boyd_DFWM3} or those using H$_2$~\cite{bib:ph2_DFWMFG} have been conducted, though motivations of the previous experiments are different substantially from that in this study. 

The rest of this paper is structured as follows: In section~\ref{sec:setup} we note properties of the p-H$_2$ molecule and describe the experimental setup. 
In section~\ref{sec:results} experimental results are described. 
In section~\ref{sec:discussion} these results are compared to the results of the numerical simulations. 
Detailed information about the derivation of the Maxwell-Bloch equations and the numerical simulation is described in Appendix.

\section{Experimental setup}\label{sec:setup}
Para-hydrogen (p-H$_2$) is hydrogen molecule whose spins of the two nuclei are aligned antiparallel and the total nuclear spin angular momentum $I_N$ is zero. 
Rotational quantum numbers of p-H$_2$ in the electronic ground state are even ($J=0,2,\dots$) because of the Pauli exclusion principle. 
At low temperature, most of p-H$_2$ molecules lie in the $J=0$ state. 

In order to achieve huge rate amplification, it is favorable to use a target with small decoherence. 
The intermolecular interaction is one of the causes of decoherence.
The intermolecular interactions of $J=0$ p-H$_2$ molecules are weaker than those of ortho-hydrogen molecules ($I_N=1$, odd-$J$) because of the spherical symmetry of rotational wavefunction of them. 
For this reason we use the p-H$_2$ target rather than normal hydrogen gas target.  

\begin{figure}
\centering
\includegraphics[width=0.4\textwidth]{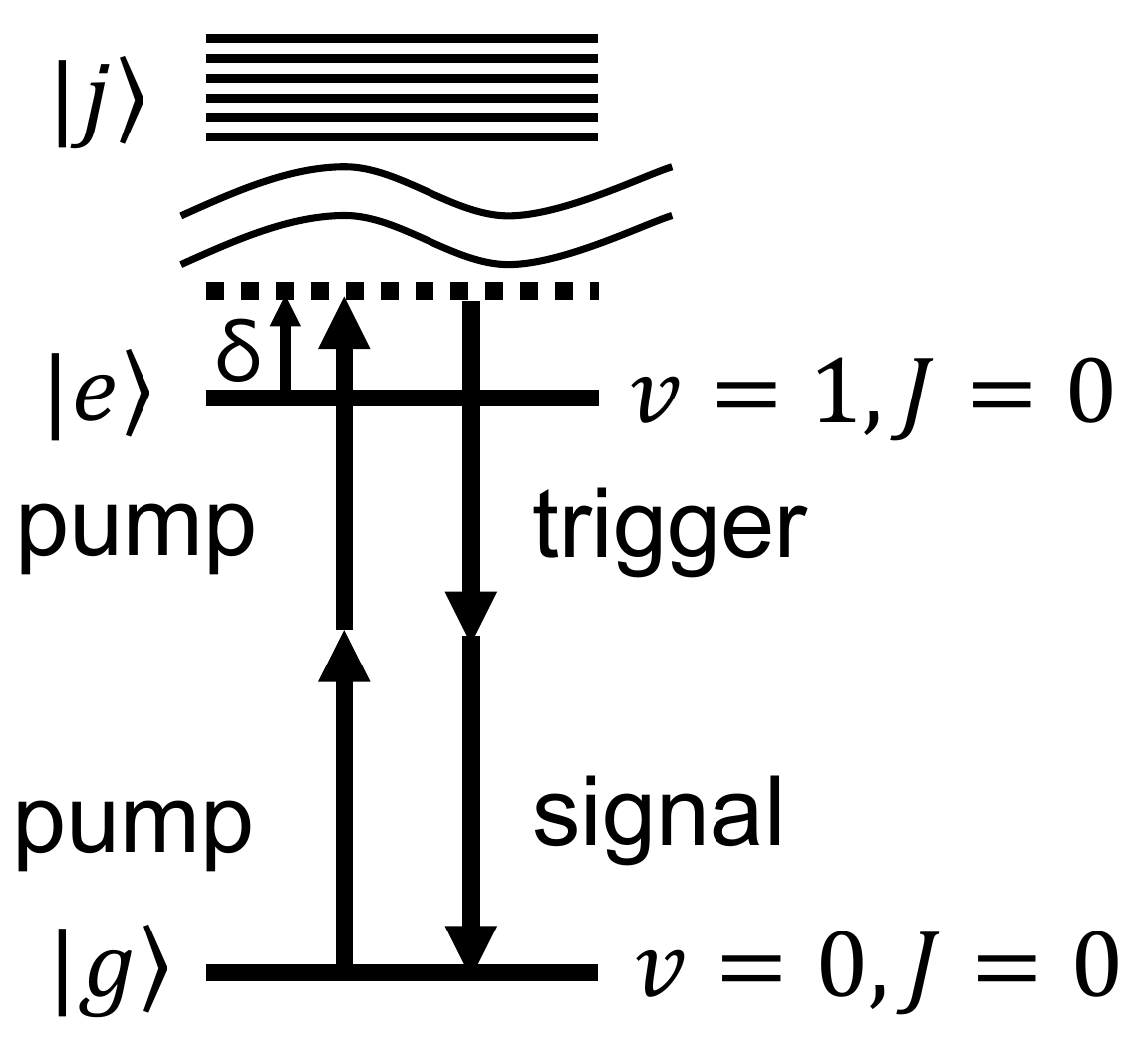}
\caption{Schematic energy-level diagram of the parahydrogen molecule. The energy difference between the $v=0$ ground state $|g\rangle$ and the $v=1$ \textcolor{black}{vibrationally excited} state $|e\rangle$ corresponds to a wavelength of 2403 nm. Intermediate states $|j\rangle$ of the two-photon transitions are electronically excited states. 
The two-photon detuning $\delta$ is defined as $\delta=2\omega_l-\omega_{eg}$.}
\label{fig:diagram}
\end{figure}

Figure~\ref{fig:diagram} shows a schematic energy-level diagram of the p-H$_2$ molecule. 
We generate coherence between the ground state $|g\rangle$ ($v=0$, $J=0$) and the first vibrationally excited state $|e\rangle$ ($v=1$, $J=0$).
The energy difference between $|g\rangle$ and $|e\rangle$ is $\hbar\omega_{eg} =$ 0.5159 eV.
Single-photon electric dipole ($E1$) transitions between these states are forbidden and two-photon $E1\times E1$ transitions are allowed. 
For this reason $|e\rangle$ is metastable and its spontaneous emission rate is $\mathcal{O}(10^{-11})$ Hz, which is dominated by the two-photon $E1\times E1$ transition process~\cite{bib:TPE1}. 
Effect of the decoherence due to the spontaneous de-excitation is negligibly small in this setup.  
Intermediate states $|j\rangle$ of the two-photon transitions are electronically excited states.
The energy level differences between $|g\rangle$ or $|e\rangle$ and $|j\rangle$, $\hbar\omega_{jg}$ and $\hbar\omega_{je}$ are much larger than $\hbar\omega_{eg}$ (far-off-resonance condition). 
In the previous coherent two-photon emission (TPE) signal generation experiments via one-side excitation~\cite{bib:TPE1,bib:TPE2,bib:TPE3}, coherence between $|g\rangle$ and $|e\rangle$ was prepared using the Raman process, where \textcolor{black}{lasers with different frequency} were necessary. 
In this experiment, in contrast, we generate coherence using identical-frequency photons which originate from a single laser.   
This laser frequency is denoted by $\omega_l$ and the two-photon detuning $\delta$ is defined as $\delta=2\omega_l-\omega_{eg}$.

\begin{figure}
\centering
\includegraphics[width=0.8\textwidth]{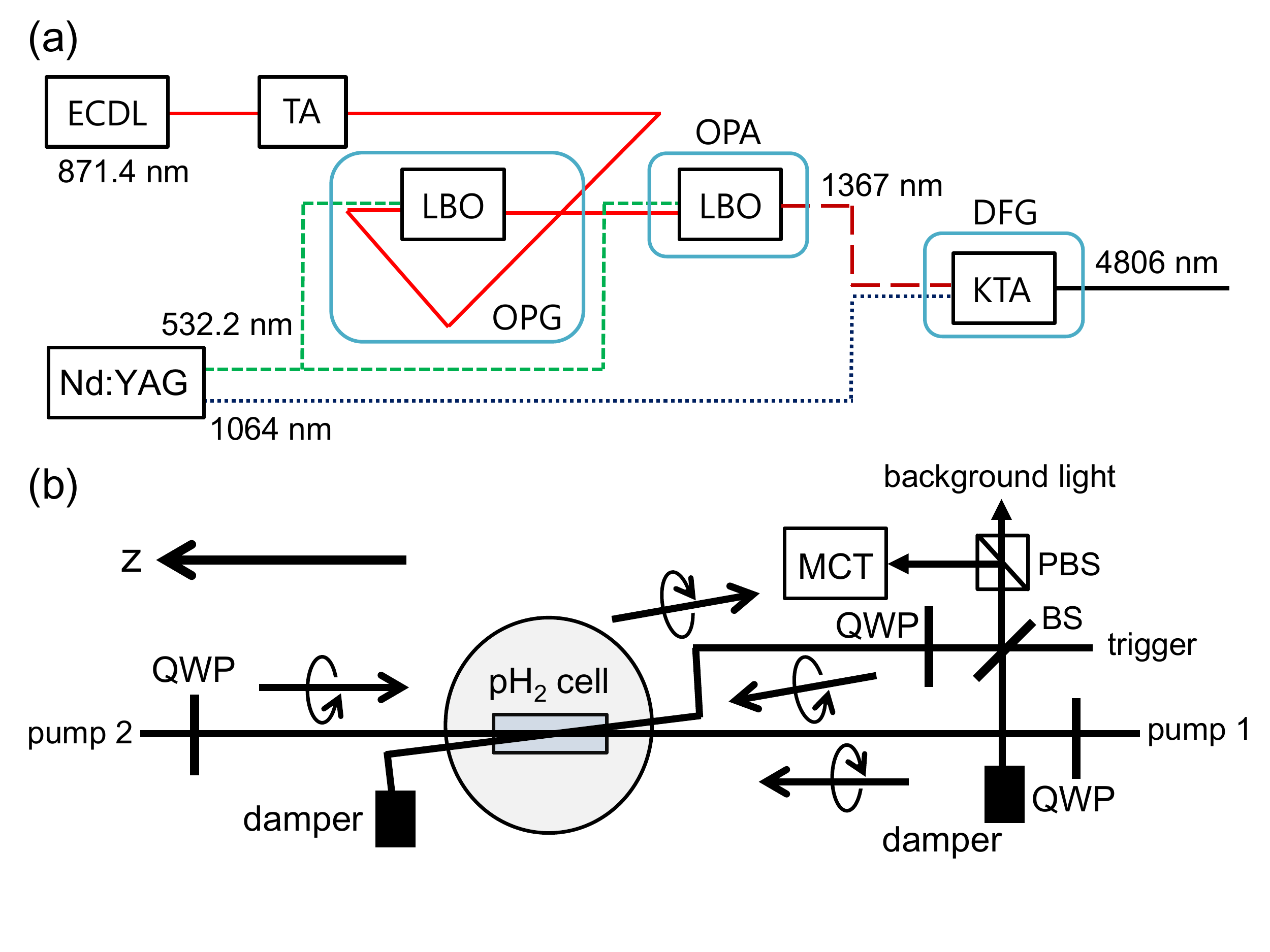}
\caption{(a) Schematic view of the laser system with the external cavity diode laser (ECDL), the tapered amplifier (TA), the lithium triborate crystals (LBO), and the potassium titanyl arsenate crystals (KTA). 
(b) Schematic view of the experimental setup with the quarter-wave plates (QWP), the beam splitter (BS), the polarizing beam splitter (PBS), and the mercury cadmium telluride (MCT) detector. The pump1 (pump2) beam was injected from the right (left) side of the figure. The trigger beam was injected from the right side of the figure.}
\label{fig:setup}
\end{figure}

The laser system is basically the same as the previous experiment where third-harmonic generation from the p-H$_2$ gas target was observed~\cite{bib:THG}.
Here we describe the laser system briefly. 
Figure~\ref{fig:setup} (a) shows the schematic view of the mid-infrared (MIR) laser system. 
A continuous-wave (cw) laser ($\lambda=871.4$ nm) was prepared using the external cavity diode laser (ECDL). 
The $\delta$ was precisely controlled by adjusting the frequency of the ECDL. 
The cw near infrared laser intensity was amplified by the tapered amplifier (TA). 
The cw laser was injected into the triangle cavity, where pulse near-infrared laser was generated in lithium triborate (LBO) crystals with the second harmonic of an injection-seeded Nd:YAG \textcolor{black}{pulsed} laser (Continuum, Powerlite DLS 9010, $\lambda=532.2$ nm) through optical parametric generation (OPG). 
Next, through optical parametric amplification (OPA) in LBO crystals, 1367 nm pulses were generated from the \textcolor{black}{pulsed} near-infrared laser and the second harmonic of the injection-seeded Nd:YAG pulse laser. 
The MIR pulses ($\lambda=4806$ nm) were then generated through difference-frequency generation (DFG) in potassium titanyl arsenate (KTA) crystals between the fundamental pulses of the injection-seeded Nd:YAG laser ($\lambda=1064$ nm) and the 1367 nm pulses. 
The repetition rate and duration of the MIR pulses were 10 Hz and roughly 5 ns (FWHM), respectively. 
Laser linewidth of the MIR pulses was measured with the absorption spectroscopy using rovibrational transitions of the carbonyl sulfide gas. Estimated FWHM linewidth is 145 (16) MHz, which is roughly 1.6 times larger than the Fourier-transform-limited linewidth if Gaussian-shaped pulses are assumed. 

Figure~\ref{fig:setup} (b) shows the schematic view of the experimental setup.
High-purity ($>99.9\%$) p-H$_2$ gas was prepared from normal hydrogen gas by using a magnetic catalyst (Fe(OH)O) cooled to approximately 14 K.
The prepared p-H$_2$ gas was enclosed in a copper cell, which was located inside a cryostat \textcolor{black}{kept} at approximately 78 K by liquid nitrogen. 
At this temperature, a conversion rate from p-H$_2$ molecules to orthohydrogen molecules during the experiment is slow and it is enough to replace the p-H$_2$ gas target roughly once a week. 
The length of the cell was 15 cm, which was much shorter than the MIR pulse length.
Anti-reflection coated BaF$_2$ (Thorlabs, WG01050-E) substrates were used for the optical windows of the cryostat and the cell. 

The MIR beam was divided into three beams by using beam splitters. 
For excitation of p-H$_2$ gas, counter-propagating pump MIR pulses (pump1, pump2) were injected into the cell.
\textcolor{black}{The third} beam, the trigger laser, was injected into the cell simultaneously with the pump beams to induce the TPE process. The timing jitter among the beams was negligibly small. 
The trigger beam was roughly 1$^{\circ}$ tilted from the pump beam axis in the horizontal direction so that the signal light was separated from the pump beams while keeping overlapped region between the pump and trigger beams.
The trigger beam was almost overlapped with the pump beams around the center of the target cell. 
Diameter (D4$\sigma$) of the input beams were $2-3$ mm and were loosely focused around the center of the cell. 
The input pulse energies of the beams were measured by using an energy detector (Gentec-EO, QE12).
Measured energies for the pump beams $\mathcal{E}_{\rm p1}$ and $\mathcal{E}_{\rm p2}$ were roughly 1 mJ/pulse and that 
for the trigger beam $\mathcal{E}_{\rm trig}$ was 0.6 mJ/pulse. 
There existed roughly a 10\% pulse-by-pulse energy fluctuations. 
 
The pump beams and the trigger beam were circularly polarized by using quarter-wave plates before they were injected into the target.  
Both of the pump lasers were right-handed circularly polarized (RH) from the point of view of the source. 
The $z$ axis is defined in figure~\ref{fig:setup} (b).
We chose the $z$ axis as the quantization axis.  
The polarization of the pump1 (pump2) beam was $\sigma^+$ ($\sigma^-$) and the excitation process by the pump1 (pump2) beam was a $\Delta m_J=+1$ ($\Delta m_J=-1$) process. 
The two-photon excitation with only the pump1 beam or the pump2 beam was forbidden because of the selection rule of the \textcolor{black}{$J=0 \to J=0$} two-photon transition. 
In contrast, the two-photon excitation by the counter-propagating photon pair was allowed. 
The signal light was generated by the trigger pulse from excited p-H$_2$ molecules and went back along with the trigger beam line because of the amplification condition (momentum conservation). 
The trigger light was left-handed circularly polarized (LH),  and because of the selection rule the signal light was also left-handed circularly polarized. 
It was actually allowed that p-H$_2$ molecules were excited by the pump1 laser and the trigger laser.
However, in this case \textcolor{black}{the} direction of the de-excited photons was the same as that of the pump1 or the trigger laser because of the amplification condition and these photons were not observed by the signal detector.
Excitation by the pump2 and the signal photons were also allowed but its effect on signal energy was  negligibly small. 

The signal light was horizontally polarized after it passed the quarter-wave plate. 
In this experiment it was difficult to reduce background component, which comprised scattering lights of the pump and the trigger beams.
This is because the wavelength of these lights were the same as that of the signal light and wavelength filters could not be used for the reduction of the background stray light. 
A portion of the signal light was reflected by using a beam splitter (Thorlabs, BSW511) and was then reflected by a polarizing beam splitter (Research Electro-Optics, Product 15625) which was used for reduction of the background RH scattering lights.
Furthermore, the cryostat was placed on a rotation stage and was rotated so that the reflected lights of the pump and the trigger beams by the optical windows \textcolor{black}{became} the minimum.   
The MIR signal pulses were detected by using a mercury-cadmium-telluride detector (Vigo system, PC-3TE-9).

\section{Results}\label{sec:results}
\subsection{Detuning curve} 
First, we confirmed the two-photon transition signal was generated from the p-H$_2$ target. 
The signal energy varied by changing the detuning $\delta$ and could be separated from the background component.   
Figure~\ref{fig:detuning} shows observed spectra as a function of $\delta$, where target gas pressure was set to 280 kPa. 
At each detuning, the signal pulse was measured 200 or 300 times. 
The error bars in the plot indicate the standard errors, where only the statistical uncertainty is considered. 
The signal energy was fluctuated because of the fluctuation of position and energies of the input beams during the experiment. 
The systematic uncertainty which includes these fluctuations is much larger than the statistical uncertainty, but  its estimation is difficult and we do not discuss further in this paper. 
For comparison, a signal energy distribution when the polarization of one of the pump lasers was changed to LH is shown in blue square points. 
In this case, the two-photon excitation process was suppressed and only background scattering light was observed. 
This confirms the signal peak stems from the TPE process. 

The solid line in figure~\ref{fig:detuning} is a fit to data points by a Lorentzian function and a constant. 
The constant term of the fit represents the background components. 
As mentioned in the previous section, the origin of the $\delta$ is defined as the point $2\omega_l=\omega_{eg}$. 
The MIR laser frequency was estimated during the experiment by monitoring the laser frequencies of the ECDL and the seed laser of the Nd:YAG laser with a high-precision wavemeter (HighFinesse WS-7). 
Before this experiment was conducted we measured the $\omega_{eg}$ through the Raman scattering process between $\ket{g}$ and $\ket{e}$. 
The previously measured $\omega_{eg}$ was $124748.7-10.9 \times p$ GHz at 78 K, where $p$ indicates the p-H$_2$ pressure in MPa and the second term represents the pressure shift. This value is consistent with those measured by external experiments~\cite{bib:linewidth, bib:energylevel, bib:pressureshift} within the systematic uncertainty of our measurement. 
The center value of the Lorentzian spectrum by the fit was -27 MHz.
The uncertainty of the $\delta$ is roughly 170 MHz, which is determined from the absolute accuracy of the wavemeter, and this center value is close to the origin. 
The width of the fitted Lorentzian profile is described in subsection~\ref{sec:pdep}.

By considering the detector responsivity and transmittance of optics, the signal energy $\mathcal{E}_{\rm sig}$ around $\delta=0$ is estimated to be roughly 20 nJ.
This value is $\mathcal{O}(10^{-5})$ times smaller than those of pump and trigger pulses. 
We have defined the ``enhancement factor'' as a ratio of the observed photon number to that expected due to spontaneous two-photon emissions with experimental acceptance~\cite{bib:TPE1,bib:TPE4}. 
The enhancement factor in the current experiment is roughly comparable to that in the previous measurement~\cite{bib:TPE2}.

\subsection{Signal energy dependence on input pulse energies}\label{sec:edep}
Next, we present a dependence of signal energy on the energies of the pump and the trigger beams. 
\textcolor{black}{To this end, we varied input pulse energies of MIR pulses by using neutral density (ND) filters. 
These ND filters were placed before MIR pulses were divided into the pump and trigger beams and the pump and trigger beam energies varied at the same rate.}

\begin{figure}
\centering
\includegraphics[width=0.7\textwidth]{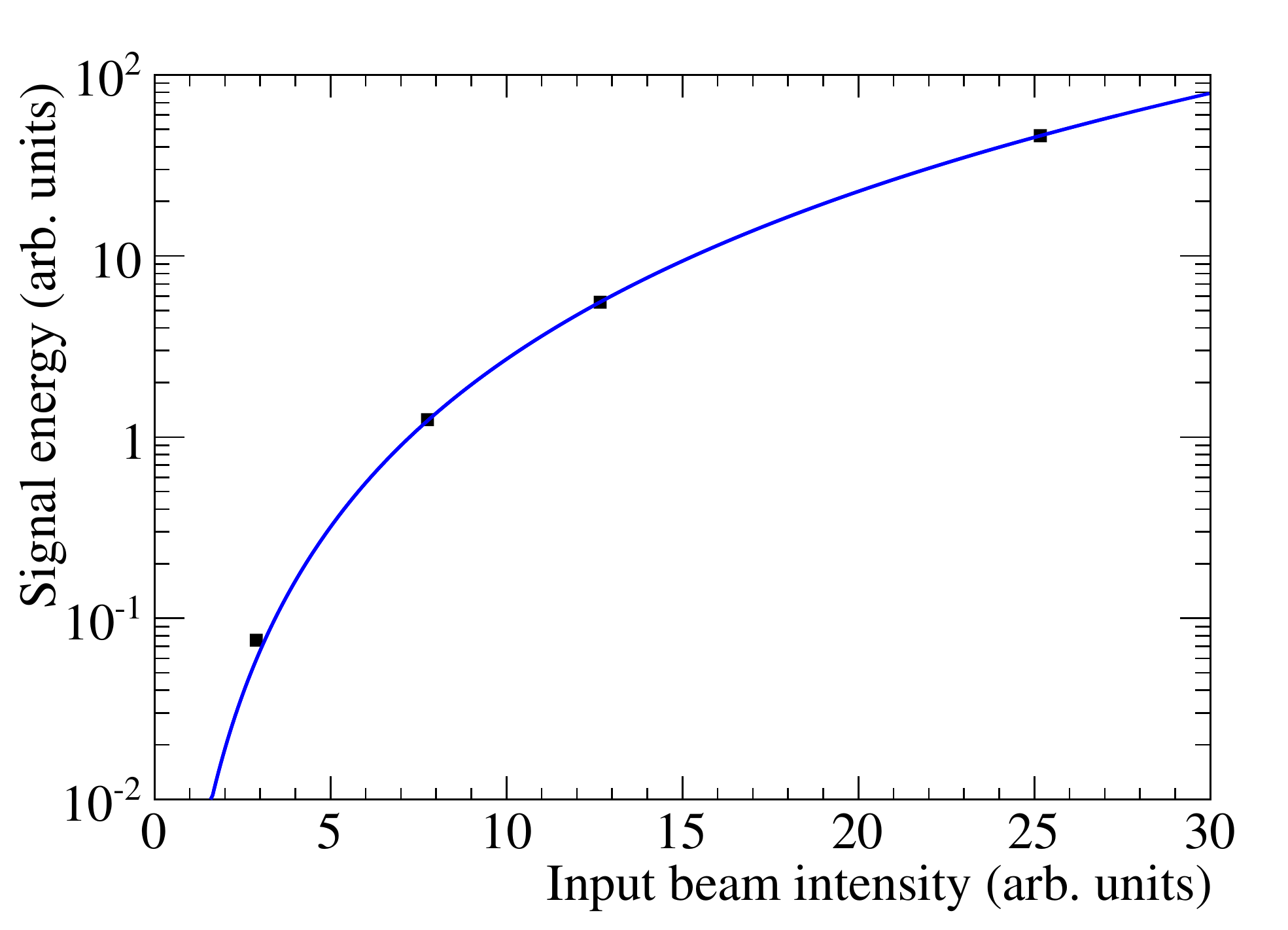}
\caption{Dependence of signal energy $\mathcal{E}_{\rm sig}$ on input beam intensity $I$. 
Black square points indicate experimental data. 
At each data point measurement was conducted approximately 3000 times and background components were subtracted by using data taken at off-resonant point.
Blue solid line indicates experimental data fitted with $\mathcal{E}_{\rm sig}=A\times I^{B}$, where $A$ and $B$ are the fit parameters. 
}
\label{fig:edep}
\end{figure}

Figure~\ref{fig:edep} shows the dependence of signal energy at 288 kPa and $\delta\approx0$ on input beam energy. 
At each data point measurement was conducted approximately 3000 times and background components were subtracted by using signal energy data measured at off-resonant points. 
Blue solid line indicates experimental data fitted with $\mathcal{E}_{\rm sig}=A\times I^{B}$, where $I$ indicates the input pulse intensity before division and $A$ and $B$ are the fit parameters.
We obtained $B=2.99\pm0.03$, though actual systematic uncertainty is considered to be much larger than the obtained error. 

\subsection{Dependence of detuning curves on target pressure}\label{sec:pdep} 
Finally, we present dependences of detuning curve width and signal energy on p-H$_2$ gas pressure.
Target p-H$_2$ gas pressure was varied from 10 to 340 kPa. 
Signal energies especially in the low-pressure region were weak and were largely affected by the fluctuation of background lights and detector noises.
For confirmation of reproducibility, measurements were conducted several times in 10-60 kPa region. 
Peak energies and detuning curve widths were obtained from fitted detuning dependence curves. 
The \textcolor{black}{fitting error sizes} were calculated \textcolor{black}{with the method of unweighted least squares}. 

\begin{figure}
\centering
\includegraphics[width=0.7\textwidth]{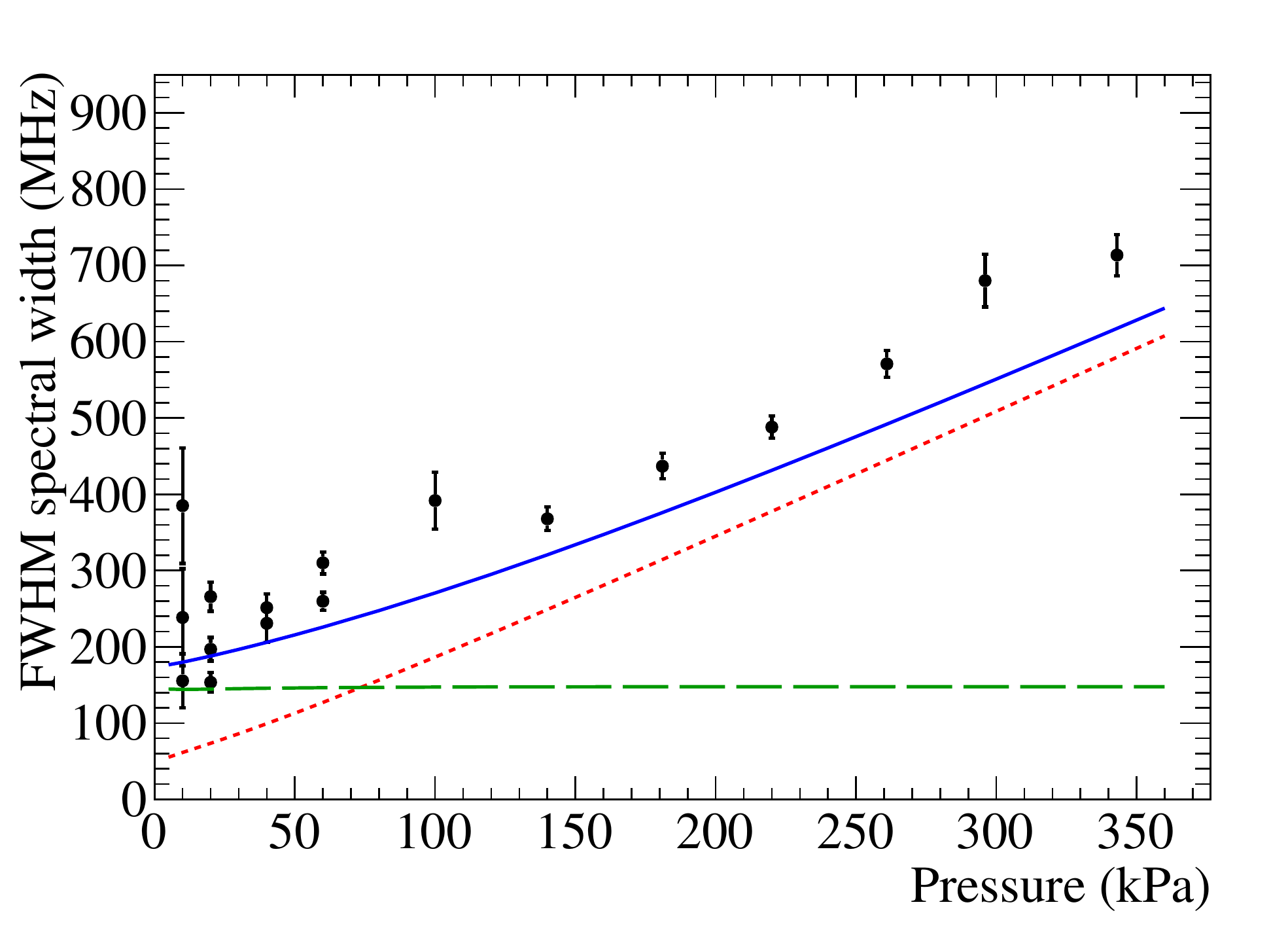}
\caption{Target pressure dependence of spectral widths of detuning curves. Black circle points indicate experimental data. 
In the low-pressure region, measurements were conducted several times for confirmation of reproducibility. 
Blue solid line indicates the FWHM width of the simulation result. 
Red dotted line and green dashed line indicate the Lorentzian component and the Gaussian component of the Voigt width, respectively.
}
\label{fig:pres_width}
\end{figure}

Dependence of the FWHM detuning curve width on target pressure is shown in figure~\ref{fig:pres_width}. 
In the low-pressure region the detuning curve width is considered to be dominated by the laser linewidth. 
The detuning curve width is increasing as target pressure becomes higher, which is due to the pressure broadening effect.
The size of the pressure broadening effect is approximately proportional to the gas pressure~\cite{bib:linewidth}. 

\begin{figure}
\centering
\includegraphics[width=0.7\textwidth]{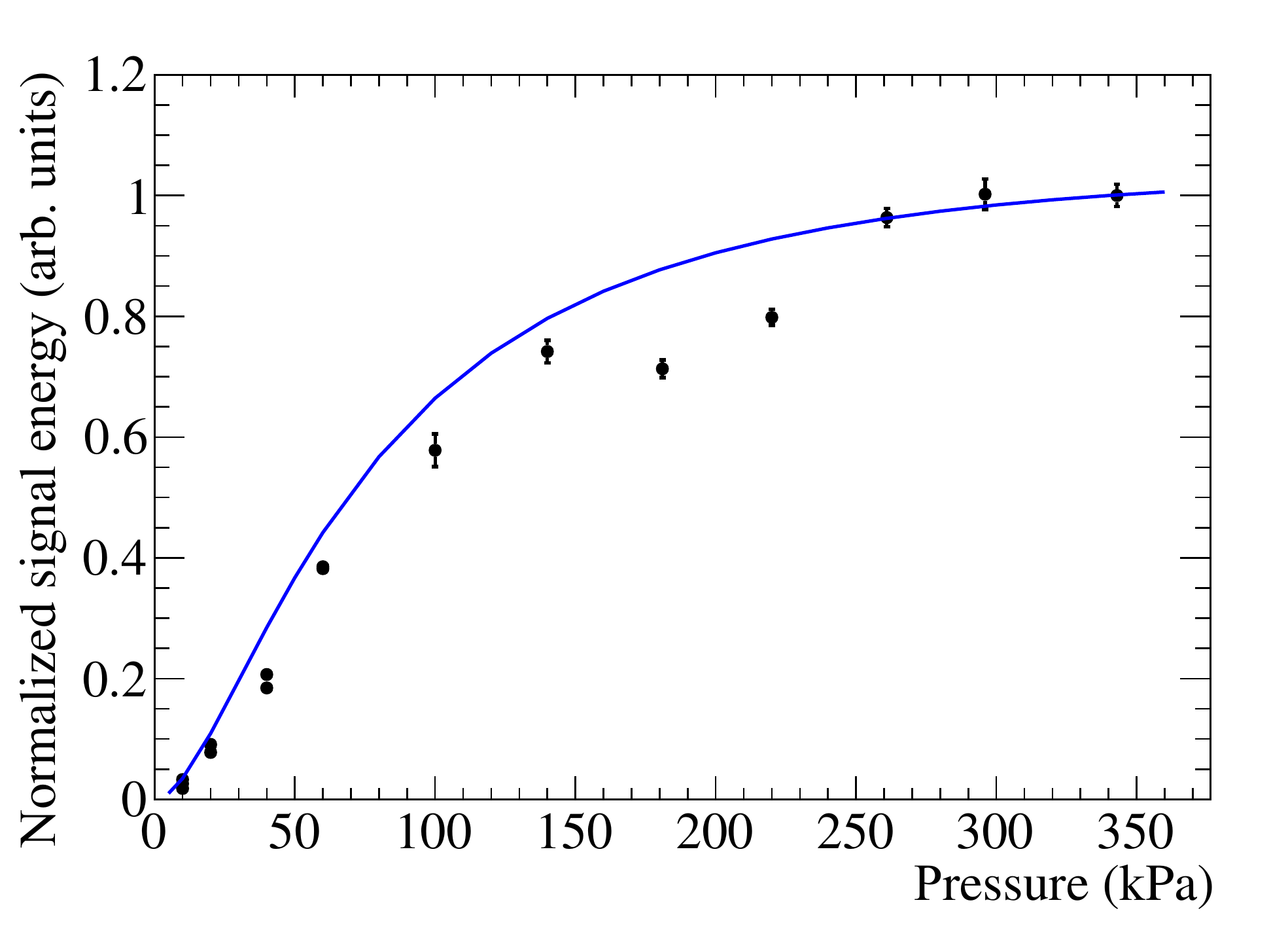}
\caption{Target pressure dependence of peak signal energy normalized by that of the measured data at the highest pressure. Black circle points indicate experimental data. In the low-pressure region, measurements were conducted several times for confirmation of reproducibility. Blue line indicates the simulation result.}
\label{fig:pres_height}
\end{figure}

Dependence of the signal peak energy on target pressure is shown in figure~\ref{fig:pres_height}. 
The data points are normalized by that at the highest pressure. 
Signal energy increases as the target density becomes higher because more p-H$_2$ molecules are excited.
On the other hand, the increasing rate of the signal energy becomes lower because effect of decoherence by the pressure broadening is larger in the high-pressure region. 

\begin{figure}
\centering
\includegraphics[width=0.7\textwidth]{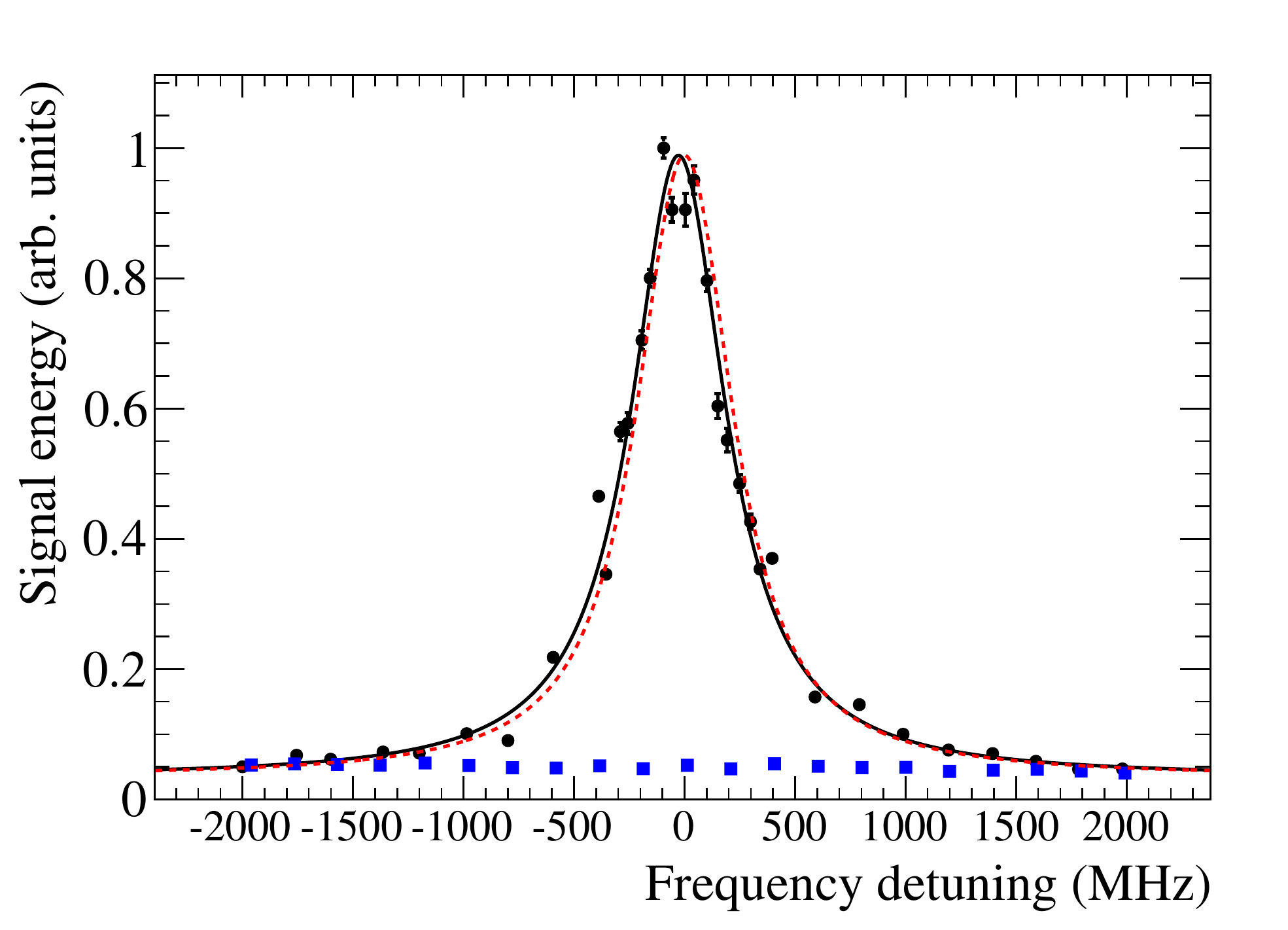}
\caption{Signal energy distributions as a function of detuning $\delta$ for RH+RH polarization pump beams (black circle points) and RH+LH polarization pump beams (blue square points). 
At each point, signal pulse was measured 200 or 300 times. The error bars indicate standard errors.
Black solid line indicates a fit to black points by a Lorentzian function and a constant. 
The constant term of the fit represents the background components. 
Red dashed line indicates the normalized detuning curve of the simulation result with the same constant background.
}
\label{fig:detuning}
\end{figure}

\section{Discussion}\label{sec:discussion}
\subsection{Maxwell-Bloch equations}\label{sec:discussion1}
In order to understand the experimental results \textcolor{black}{qualitatively}, we constructed a numerical simulation which reproduces the experimental situation. 
The simulation is based on Maxwell-Bloch equations with one spatial and one temporal (1+1)  dimension. 
In this simulation, intensity and duration of the pump beams and the trigger beam, and p-H$_2$ gas target pressure were set to the same values as those of the experiment. 
The temporal shape of the input pulses are assumed to be Gaussian. 
Temporal electric field distribution of each input beam at the target edge is written as 
\begin{equation}
|E_{\rm X}(t)|^2=\frac{2c}{\varepsilon_0\sqrt{2\pi}\sigma_t}I_{\rm X}\exp\left(-\frac{(t-t_0+L/2c)^2}{2\sigma_t^2}\right),
\end{equation}
where $I_{\rm X}$ (X=p1, p2, trig), $\sigma_t$, and $L$ indicate pulse intensity of each input beam, duration of each pulse, and the target length, respectively. The center of these beams passes through the target center at $t=t_0$.
We describe the whole equations in Appendix and here we describe an essential part of the equations:
\begin{eqnarray}
\left(\frac{\partial}{\partial t}+c\frac{\partial}{\partial z}\right) E_{\rm sig} \approx \frac{{\rm i}\omega_lN_t}{2}\left(\alpha_{gg}E_{\rm sig}+2\alpha_{eg}\rho_{ge}^{0*}E_{\rm trig}^{*}\right), \label{eq:Maxwells4_approx} \\ 
\frac{\partial\rho_{ge}^{0}}{\partial t} = -({\rm i}\delta+\gamma_{2})\rho_{ge}^{0} -{\rm i}\Omega_{eg}^{0*}, \label{eq:Blochs3_0} \\
\Omega_{eg}^{0}\approx \frac{\varepsilon_0\alpha_{eg}}{2\hbar}E_{\rm p1}E_{\rm p2}. \label{eq:Rabiapprox}
\end{eqnarray}
Equation (\ref{eq:Maxwells4_approx}) represents the development of the envelope of the signal electric field $E_{\rm sig}$.
$N_t$ indicates the number density of the p-H$_2$ and $\alpha_{gg}$ and $\alpha_{eg}$ indicate polarizabilities of the p-H$_2$. 
Equation (\ref{eq:Blochs3_0}) represents the temporal development of the coherence generated by the pump lasers, where $\gamma_2$ indicates the transverse relaxation rate. The $\Omega_{eg}^{0}$ represents the two-photon Rabi frequency. 

In the previous one-side laser injection scheme, the Maxwellian part of the partial differential equations 
can be simplified to ordinary differential equations by introducing the co-moving coordinates~\cite{bib:TPE1}. 
\textcolor{black}{However, this treatment cannot be used in the case of the counter-propagating injection scheme.}
\textcolor{black}{In this paper, }we directly treat the partial differential equations by using the method of lines~\cite{bib:methodoflines}
After that treatment we numerically solve ordinary differential equations where the space derivative ($\partial$/$\partial z$) is discretized. 
For the discretization of the space derivative, we used the weighted essentially non-oscillatory schemes~\cite{bib:WENO}. 

In this simulation the $\gamma_2$ is an input parameter. 
In previous one-side excitation experiments~\cite{bib:ptepreview} we used the Raman transition linewidth $\Delta_{\mathrm{Raman}}$ between $|g\rangle$ and $|e\rangle$ as the value of the $2\gamma_2$. 
The pressure dependence of $\Delta_{\mathrm{Raman}}$ is approximately written~\cite{bib:linewidth} as
\begin{equation}
\Delta_{\mathrm{Raman}}=C_p p + \frac{C_{\rm D}}{p}, \label{eq:Ramanwidth} 
\end{equation} 
where $p$ indicates the p-H$_2$ pressure and $C_p$ and $C_{\rm D}$ are constant.
The first term represents the pressure broadening effect and the second term represents the Doppler broadening effect with Dicke narrowing~\cite{bib:Dicke}.
In the current experimental scheme, in contrast, the signal photons were mostly generated from $|e\rangle$ excited by the same-frequency counter-propagating pump lasers. 
Thus the system is almost Doppler-free~\cite{bib:Dopplerfree1,bib:Dopplerfree2} and the size of the residual Doppler effect is negligibly small compared to the other broadening effect. 
For this reason we take into account only the pressure broadening term as the $\gamma_2$. 

From equations (\ref{eq:Maxwells4_approx})-(\ref{eq:Rabiapprox}), it is found that signal energy depends on input pulse energies, the $\delta$, $N_t$, and the $\gamma_2$. 
We can confirm consistency of signal energy dependences of these parameters between the measured data and the simulation results from the detuning curves and the signal energy dependences on input pulse energy and target pressure. 

\subsection{Analytical estimation of the signal energy dependence on input pulse energies}\label{sec:discussion2}
We can predict how the input energy dependence behaves from the Maxwell-Bloch equations in a simplified case where we assume the energies of the pump and the trigger beams are constant.
They actually decrease due to the two-photon absorption, but their decrease rates after the beams pass through the target are expected to be $\mathcal{O}$(1)\%.
\textcolor{black}{Thus this treatment is a reasonable approximation.}  
Temporal development of the coherence (equation~(\ref{eq:Blochs3_0})) is \textcolor{black}{written as}  
\begin{equation}
\frac{\partial\rho_{ge}^{0}}{\partial t} = -({\rm i}\delta+\gamma_{2})\rho_{ge}^{0}-{\rm i}\frac{c\alpha_{eg}}{\hbar\sigma_t\sqrt{2\pi}}\sqrt{I_{\rm p1}I_{\rm p2}}\exp\left(-\frac{(t-t_0)^2}{2\sigma_t^2}-\frac{z^2}{2\sigma_t^2c^2}\right), 
\label{eq:approx_coh_diff}
\end{equation}
where $\sigma_t$ indicates duration of each pulse. 
By considering the initial condition $\rho_{ge}^{0}=0$ at $t=-\infty$, 
\textcolor{black}{we obtain the solution of equation (\ref{eq:approx_coh_diff}) as}
\begin{eqnarray}
\rho_{ge} &=& -{\rm i}\frac{c\alpha_{eg}}{2\hbar}\sqrt{I_{\rm p1}I_{\rm p2}}\exp\left(-\Gamma(t-t_0)+\frac{\sigma_t^2\Gamma^2}{2}-\frac{z^2}{2\sigma_t^2c^2}\right) \nonumber \\
&\times& {\rm erfc}\left(\frac{-(t-t_0)+\sigma_t^2\Gamma}{\sqrt{2}\sigma_t}\right),
\label{eq:approx_coh}
\end{eqnarray}
where $\Gamma=\gamma_{2}+\rm{i}\delta$ and erfc indicates the complementary error function.
From equation (\ref{eq:approx_coh}) $|\rho_{ge}^{0}|$ is proportional to $\sqrt{I_{\rm p1}I_{\rm p2}}$.
In the development of the signal electric field (equation~(\ref{eq:Maxwells4_approx})), the last term of the right-hand side indicates generation of the signal photons induced by the trigger field. 
Since $I_{\rm sig}\propto|E_{\rm sig}|^2$ is proportional to 
\begin{equation}
N_t^2|\rho_{ge}^0|^2I_{\rm trig}\propto N_t^2I_{\rm p1}I_{\rm p2}I_{\rm trig}, \label{eq:approx}
\end{equation}
the signal energy $\mathcal{E}_{\rm sig}$ has the cubic dependence on $I$. 
We also obtained $\mathcal{E}_{\rm sig}=AI^{x}$, $x\approx 3.0$ from the simulation, where this approximation is not applied.
The measured data are consistent with the theoretical predictions and the simulation result. 

\subsection{Comparison of the detuning curve width dependence on target pressure}\label{sec:discussion3}
The detuning curve shape of the simulation is well-approximated by a Voigt profile, which is the convolution of a Lorentzian profile and a Gaussian profile. 
Red dashed line in figure~\ref{fig:detuning} shows the signal energy dependence on the detuning obtained by
the simulations.
Here the signal energy is normalized so that the peak height of the detuning curve of the simulation is the same as that of the fitted data. 
The detuning curve shape of the simulation result is consistent with that of the fitted data.  

Blue solid line in figure~\ref{fig:pres_width} indicates detuning curve widths calculated from the simulations. 
The detuning curve width of the experimental data is slightly wider with the simulation results. 
A width of a Voigt profile\textcolor{black}{, $w_{\rm V}$,} is approximately written as $w_{\rm V}\approx0.5346w_{\rm L} + \sqrt{0.2166w^2_{\rm L} + w^2_{\rm G}}$, where $w_{\rm L}$ (red dotted line), and $w_{\rm G}$ (green dashed line) indicate the FWHM width of Lorentzian and Gaussian function, respectively~\cite{bib:Voigt}. 
In this case $w_{\rm L}$ is almost linear to the p-H$_2$ pressure and $w_{\rm G}$ is almost constant. 
The Gaussian term arises from the Fourier-transform-limited laser linewidth.
As mentioned in section~\ref{sec:setup}, actual laser linewidth is wider than the Fourier-transform-limited one.
This is considered to be the reason of the difference of the detuning curve width between the data and the simulation. 

Blue line in figure~\ref{fig:pres_height} indicates the pressure dependence of the signal peak energy obtained by the simulation. 
Signal energy of the simulation is also normalized by that of the data point at the highest pressure.
Pressure dependences of normalized signal energy of the simulation are also basically consistent with the data. 
These results indicate decoherence of the system is dominated by the pressure broadening effect. 
There exists a dip near the pressure of 200 kPa in the data. 
This is considered to be due to unexpected variation of some of beam properties during taking those data points. 

\subsection{Comparison of absolute signal energy}\label{sec:discussion4}

Finally, we mention a comparison of the absolute signal energy between the simulation result and the measured data. In the calculation of the signal energy of the simulation we assume the spatial profile of the input beams are assumed to be flat-top and have a diameter of 2 mm. 
We calculate the absolute signal energy by integrating the temporal profile of the signal pulse obtained by the numerical simulation. 
The signal energy at $\delta=0$ and at 280 kPa of the simulation is roughly 2.2$\times10^2$ nJ, which is roughly 10 times larger than that of the data. 

Signal energies largely depend on the size of the relaxation rate. 
However, as described in the previous subsection, normalized signal energy dependence on target pressure of the simulations is generally consistent with those of the measured data. 
Thus it is unlikely that wrong estimation of the relaxation rate is the cause of the discrepancy. 
\textcolor{black}{One possible} reason for this discrepancy is that our numerical simulation is too simplified.
In particular, in this simulation we assume that the trigger and the pump beams are completely overlapped throughout the target, which is not correct in the actual experiment because the trigger beam was tilted from the pump beamline. 
By considering the overlapped region of the beams inside the target, the generated signal energy with this effect is estimated to be roughly half compared with the case where the beams completely overlap. 
However, the absolute signal energy of the simulation is still roughly five times larger than that of the data.
Another simplification is that we numerically solved $1+1$-dimensional Maxwell-Bloch equations, where transverse terms of the laser fields were not taken into account. 
If we execute more realistic simulations, this discrepancy might be resolved, though this is beyond the scope of the current paper. 

\section{Conclusion}\label{sec:conclusion}
In order to deepen understandings of the rate amplification mechanism using atomic or molecular coherence, we conducted an experiment where coherence was prepared by counter-propagating excitation laser pulses. 
We successfully observed the two-photon emission (TPE) signal from parahydrogen (p-H$_2$) molecules. 
This observation is an important step for future neutrino spectroscopy experiments because the counter-propagating excitation scheme will be adopted in those experiments, though further studies are necessary for observation of RENP processes. 

It is also an important advance to improve understandings of the rate amplification mechanism through numerical simulations. 
We numerically solved Maxwell-Bloch equations with one spatial and one temporal dimension.  
We compared the signal energies of the measured data and those dependences on p-H$_2$ gas pressure and input beam energies with those calculated by using the numerical simulations. 
We found dependences of the data are qualitatively consistent with the simulation.

\ack
This work was supported by JSPS KAKENHI Grant No. JP15H02093, No. JP15H03660, No. JP15K13486, No.
JP15K17651, No. JP16J10938, No. JP17K14292, No. JP17K14363, No. JP17K18779, \textcolor{black}{No. JP17H02895}, and \textcolor{black}{No. JP17H02896}. This work was also supported by JST, PRESTO and the Matsuo Foundation.

\appendix
\section{Construction of simulation based on Maxwell-Bloch equations}\label{sec:simulation}

In this Appendix a derivation of Maxwell-Bloch equations and description of the numerical simulation are shown. 
\subsection{Laser field and the parahydrogen states of the experimental system}
The electric fields of the pump, trigger, signal lasers are represented as $\tilde{\bm{E}}_{\rm p1}$,  $\tilde{\bm{E}}_{\rm p2}$, $\tilde{\bm{E}}_{\rm trig}$, and $\tilde{\bm{E}}_{\rm sig}$, respectively. 
We assume the pump beams are right-handed circularly polarized (RH) and the trigger and the signal beams are 
left-handed circularly polarized (LH). 
In order to make the simulation simple and the numerical simulation fast, we conducted $1+1$-dimensional simulations. 
We also ignore third or higher harmonic generation processes, which are mainly generated through the two-photon excitation between the pump1 and the trigger beam. 
Our interest lies in the slowly varying envelopes of the electromagnetic fields.
Because all the fields oscillate with the frequency around the laser frequency $\omega_l$ thanks to the simplification, they are expressed 
with the electromagnetic field envelopes ($E_{\rm p1}$, $E_{\rm p2}$, $E_{\rm trig}$, and $E_{\rm sig}$) by 
\begin{eqnarray}
\tilde{\bm{E}}_{\rm p1}(z,t)&=&\frac{1}{2}\left(E_{\rm p1}(z,t)\hat{\bm{\epsilon}}_{R}\exp\Bigl({-{\rm i}\omega_l(t-\frac{z}{c})\Bigr)}+({\rm c.c.})\right), \\
\tilde{\bm{E}}_{\rm p2}(z,t)&=&\frac{1}{2}\left(E_{\rm p2}(z,t)\hat{\bm{\epsilon}}_{R}\exp\Bigl({-{\rm i}\omega_l(t+\frac{z}{c})\Bigr)}+({\rm c.c.})\right), \\
\tilde{\bm{E}}_{\rm trig}(z,t)&=&\frac{1}{2}\left(E_{\rm trig}(z,t)\hat{\bm{\epsilon}}_{L}\exp\Bigl({-{\rm i}\omega_l(t-\frac{z}{c})\Bigr)}+({\rm c.c.})\right), \\
\tilde{\bm{E}}_{\rm sig}(z,t)&=&\frac{1}{2}\left(E_{\rm sig}(z,t)\hat{\bm{\epsilon}}_{L}\exp\Bigl({-{\rm i}\omega_l(t+\frac{z}{c})\Bigr)}+({\rm c.c.})\right),
\end{eqnarray}
where $\hat{\bm{\epsilon}}_{R}$ and $\hat{\bm{\epsilon}}_{L}(=\hat{\bm{\epsilon}}_{R}^{*})$ represent circular polarization unit vectors and $\tilde{\bm{E}}$ or $\tilde{E}$ represent that electric fields include fast oscillating phase terms. 
Next, we denote the wave function of the p-H$_2$ system by 
\begin{equation}
\ket{\psi}=c_g e^{-{\rm i} \omega_g t}\ket{g}+c_e e^{-{\rm i} (\omega_e+\delta) t}\ket{e}+c_{j+} e^{-{\rm i} \omega_{j} t}\ket{j_+} + c_{j-} e^{-{\rm i} \omega_{j} t}\ket{j_-},
\end{equation}
where $\ket{j_+}$ ($\ket{j_-}$) represents $m_J=+1$ ($m_J=-1$) intermediate states of the p-H$_2$.
The Schr{\"o}dinger equation of the system is 
\begin{equation}
{\rm i}\hbar\frac{\partial}{\partial t}\ket{\psi}=(H_0+H_I)\ket{\psi},
\label{eq:schrodinger}
\end{equation}
where $H_0$ indicates the free term of the p-H$_2$ states and $H_I$ indicates the interaction Hamiltonian:
\begin{eqnarray}
H_0\ket{g}=\hbar\omega_g\ket{g},\; H_0\ket{e}=\hbar\omega_e\ket{e}, \;H_0\ket{j_\pm}=\hbar\omega_j\ket{j_\pm}, \\
H_I=-{\bm{d}}\cdot\tilde{\bm{E}}=-{\bm{d}}\cdot(\tilde{\bm{E}}_{\rm p1}+\tilde{\bm{E}}_{\rm p2}+\tilde{\bm{E}}_{\rm trig}+\tilde{\bm{E}}_{\rm sig}).
\end{eqnarray}
The $H_I$ consists of electric dipole interactions between $\ket{j_\pm}$ and $\ket{g}$ or $\ket{e}$. 
We also introduce transition electric dipole moments
$d_{jg}=\bra{j_{+(-)}}-\bm{d}\cdot\bm{\hat{\epsilon}}_{R(L)}\ket{g}$ and $d_{je}=\bra{j_{+(-)}}-\bm{d}\cdot\bm{\hat{\epsilon}}_{R(L)}\ket{e}$. 
The other transition electric dipole moments are zero since they are $E1$-forbidden transitions.
 
\subsection{Optical Bloch equations} 
In this subsection we derive Optical Bloch equations for the two-level-reduced system and their approximation for the numerical simulation. 
First, we focus on $\ket{j_\pm}$ of the Schr{\"o}dinger equation (\ref{eq:schrodinger}):
\begin{eqnarray}
{\rm i}\hbar\frac{\partial c_{j+}}{\partial t}&=&\frac{1}{2}\left(d_{jg}\exp({\rm i}\omega_{jg}t)c_g+d_{je}\exp({\rm i}\omega_{je'}t)c_e\right)\left(\tilde{E}_{\rm p1}+\tilde{E}_{\rm p2}^{*}+\tilde{E}_{\rm trig}^{*}+\tilde{E}_{\rm sig}\right), \nonumber \\
{\rm i}\hbar\frac{\partial c_{j-}}{\partial t}&=&\frac{1}{2}\left(d_{jg}\exp({\rm i}\omega_{jg}t)c_g+d_{je}\exp({\rm i}\omega_{je'}t)c_e\right)\left(\tilde{E}_{\rm p1}^{*}+\tilde{E}_{\rm p2}+\tilde{E}_{\rm trig}+\tilde{E}_{\rm sig}^{*}\right), \nonumber
\end{eqnarray}
where we introduce $\omega_{e'}=\omega_e+\delta$ and $\omega_{je'}=\omega_j-\omega_e-\delta$.
By using the Markovian approximation and an initial condition ($c_{j\pm}(t=0)=0$), we obtain 
\begin{eqnarray}
c_{j+}=-\frac{1}{2\hbar}\sum_m^{g, e'}d_{jm}c_m&&\Big(\frac{\exp({\rm i}(\omega_{jm}-\omega_l)t)-1}{\omega_{jm}-\omega_l}(\bar{E}_{\rm p1}+\bar{E}_{\rm sig}) \nonumber \\
&&+\frac{\exp({\rm i}(\omega_{jm}+\omega_l)t)-1}{\omega_{jm}+\omega_l}(\bar{E}_{\rm p2}^{*}+\bar{E}_{\rm trig}^{*})\Big), \label{eq:cjplus} \\ 
c_{j-}=-\frac{1}{2\hbar}\sum_m^{g, e'}d_{jm}c_m&&\Big(\frac{\exp({\rm i}(\omega_{jm}-\omega_l)t)-1}{\omega_{jm}-\omega_l}(\bar{E}_{\rm p2}+\bar{E}_{\rm trig}) \nonumber \\
&&+\frac{\exp({\rm i}(\omega_{jm}+\omega_l)t)-1}{\omega_{jm}+\omega_l}(\bar{E}_{\rm p1}^{*}+\bar{E}_{\rm sig}^{*})\Big), \label{eq:cjminus}
\end{eqnarray}
where $d_{je'}=d_{je}$ and $c_{e'}=c_{e}$ and
\begin{eqnarray}
\bar{E}_{\rm p1}=E_{\rm p1}\exp\Bigl({\rm i}\omega_l\frac{z}{c}\Bigr), \ \ \bar{E}_{\rm p2}=E_{\rm p2}\exp\Bigl(-{\rm i}\omega_l\frac{z}{c}\Bigr), \\
\bar{E}_{\rm trig}=E_{\rm trig}\exp\Bigl({\rm i}\omega_l\frac{z}{c}\Bigr), \ \ \bar{E}_{\rm sig}=E_{\rm sig}\exp\Bigl(-{\rm i}\omega_l\frac{z}{c}\Bigr).  
\end{eqnarray}
By using equations (\ref{eq:cjplus}) and (\ref{eq:cjminus}) and the slowly varying envelope approximation, the original Schr{\"o}dinger equation (\ref{eq:schrodinger}) is reduced to a Schr{\"o}dinger equation of the two-level system, $\ket{g}$ and $\ket{e}$:
\begin{eqnarray}
{\rm i}\hbar\frac{\partial c_{g}}{\partial t}&=&-\frac{\varepsilon_0\alpha_{gg}}{4}c_g \left(|\bar{E}_{\rm p1}+\bar{E}_{\rm sig}|^2+|\bar{E}_{\rm p2}+\bar{E}_{\rm trig}|^2\right) \nonumber \\ &&-\frac{\varepsilon_0\alpha_{eg}}{2}c_e(\bar{E}_{\rm p1}^{*}+\bar{E}_{\rm sig}^{*})(\bar{E}_{\rm p2}^{*}+\bar{E}_{\rm trig}^{*}), \label{eq:cg}\\
{\rm i}\hbar\frac{\partial c_{e}}{\partial t}&=&-\hbar\delta c_e-\frac{\varepsilon_0\alpha_{ee}}{4}c_g \left(|\bar{E}_{\rm p1}+\bar{E}_{\rm sig}|^2+|\bar{E}_{\rm p2}+\bar{E}_{\rm trig}|^2\right) \nonumber \\ &&-\frac{\varepsilon_0\alpha_{eg}^{*}}{2}c_e(\bar{E}_{\rm p1}+\bar{E}_{\rm sig})(\bar{E}_{\rm p2}+\bar{E}_{\rm trig}), \label{eq:ce}
\end{eqnarray}
where $\alpha_{gg}$, $\alpha_{ee}$, and $\alpha_{eg}$ represent polarizabilities of hydrogen.
They are given by 
\begin{eqnarray}
\alpha_{gg}(\omega)=\sum_{j}\frac{|d_{gj}|^2}{\varepsilon_0\hbar}\left(\frac{1}{\omega_{jg}-\omega}+\frac{1}{\omega_{jg}+\omega}\right), \\
\alpha_{ee}(\omega)=\sum_{j}\frac{|d_{je}|^2}{\varepsilon_0\hbar}\left(\frac{1}{\omega_{je'}-\omega}+\frac{1}{\omega_{je'}+\omega}\right), \\
\alpha_{eg}(\omega)=\sum_{j}\frac{d_{gj}d_{je}}{\varepsilon_0\hbar}\frac{1}{\omega_{je'}+\omega}. 
\end{eqnarray}
Note that addition of angular momenta should be considered in the products of the electric dipole moments. 
We use an abbreviated notation $\alpha_{gg, ee, eg}(\omega_l)=\alpha_{gg, ee, eg}$ for the case of $\omega=\omega_l$. 
In this simulation we took into account most part of the transitions via intermediate states, that is, the $0-36$th vibrational transitions of the Lyman band and $0-13$th transitions of the Werner band~\cite{bib:sqrt3}. 

The effective Hamiltonian of the two-level system is summarized by 
\begin{equation}
{\rm i}\hbar\frac{\partial}{\partial t}\left(
\begin{array}{c}
     c_g \\
     c_e
    \end{array}
\right)=H_{\rm eff}\left(
\begin{array}{c}
     c_g \\
     c_e
    \end{array}
\right), \ \ H_{\rm eff}=-\hbar\left(
    \begin{array}{cc}
     \Omega_{gg} & \Omega_{ge} \\
     \Omega_{eg} & \Omega_{ee}+\delta
    \end{array}
  \right),
\end{equation}
where $\Omega_{ee}$, $\Omega_{gg}$ represent the ac Stark shifts and $\Omega_{eg}$ represents the complex two-photon Rabi frequencies, respectively.
The $\Omega_{ee}$, $\Omega_{gg}$, and $\Omega_{eg}$ are given by
\begin{eqnarray}
\Omega_{gg}&=&\frac{\varepsilon_0\alpha_{gg}}{4\hbar}\left(|\bar{E}_{\rm p1}+\bar{E}_{\rm sig}|^2+|\bar{E}_{\rm p2}+\bar{E}_{\rm trig}|^2\right), \label{eq:acstark} \\
\Omega_{ee}&=&\frac{\varepsilon_0\alpha_{ee}}{4\hbar}\left(|\bar{E}_{\rm p1}+\bar{E}_{\rm sig}|^2+|\bar{E}_{\rm p2}+\bar{E}_{\rm trig}|^2\right), \label{eq:acstark2} \\
\Omega_{eg}&=&\Omega_{ge}^{*}= \frac{\varepsilon_0\alpha_{eg}}{2\hbar}(\bar{E}_{\rm p1}+\bar{E}_{\rm sig})(\bar{E}_{\rm p2}+\bar{E}_{\rm trig}). \label{eq:Omega1} 
\end{eqnarray} 
In order to take into account relaxation effects, we introduce the two-level density matrix by 
\begin{equation}
\rho(z,t) = \left(
    \begin{array}{cc}
      |c_g|^2 & c_gc_e^*  \\
       c_ec_g^* & |c_e|^2 
    \end{array}
  \right) = \left(
	\begin{array}{cc}
      \rho_{gg} & \rho_{ge}  \\
      \rho_{eg} & \rho_{ee} 
    \end{array}
  \right),
\end{equation}
where $|\rho_{ge}|$ represents the size of the coherence. 
Time development of the density matrix (von Neumann equation or optical Bloch equations) which includes the relaxation effect is given by
\begin{eqnarray}
 \frac{\partial\rho_{gg}}{\partial t} &=& {\rm i}(\Omega_{ge}\rho_{eg}-\Omega_{eg}\rho_{ge}) + \gamma_{1}\rho_{ee}, 
\label{eq:Bloch1} \\
 \frac{\partial\rho_{ee}}{\partial t} &=& {\rm i}(\Omega_{eg}\rho_{ge}-\Omega_{ge}\rho_{eg}) - \gamma_{1}\rho_{ee}
=-\frac{\partial\rho_{gg}}{\partial t}, 
\label{eq:Bloch2} \\
 \frac{\partial\rho_{ge}}{\partial t} &=& {\rm i}(\Omega_{gg}-\Omega_{ee}-\delta)\rho_{ge} +{\rm i}\Omega_{ge}(\rho_{ee}-\rho_{gg}) - \gamma_{2}\rho_{ge}.
\label{eq:Bloch3}
\end{eqnarray}
Parameters $\gamma_{1}$ and $\gamma_{2}$ indicate longitudinal and transverse relaxation rates respectively and represent the sizes of decoherence effect. 
In the case of the current setup, the longitudinal relaxation comprises the natural lifetime of the excited state and the transit-time broadening, but the $\gamma_{1}$ is negligibly small compared to the laser duration. 
In contrast, the $\gamma_{2}$ largely affects the signal energy.

In the Bloch equations, $\Omega_{gg}$, $\Omega_{ee}$, and $\Omega_{eg}$ include fast oscillating phases:
\begin{eqnarray}
|\bar{E}_{\rm p1}+\bar{E}_{\rm sig}|^2+|\bar{E}_{\rm p2}+\bar{E}_{\rm trig}|^2 = |E_{\rm p1}|^2 + |E_{\rm p2}|^2 + |E_{\rm trig}|^2 + |E_{\rm sig}|^2 \nonumber \\ 
+ \left(E_{\rm p1}E_{\rm sig}^{*}\exp\Bigl(2{\rm i}\omega_l\frac{z}{c}\Bigr)
+ E_{\rm p2}E_{\rm trig}^{*}\exp\Bigl(-2{\rm i}\omega_l\frac{z}{c}\Bigr)+({\rm c.c.})\right), \\
(\bar{E}_{\rm p1}+\bar{E}_{\rm sig})(\bar{E}_{\rm p2}+\bar{E}_{\rm trig}) = E_{\rm p1}E_{\rm p2}+E_{\rm trig}E_{\rm sig} \nonumber \\ +E_{\rm p1}E_{\rm trig}\exp\Bigl(2{\rm i}\omega_l\frac{z}{c}\Bigr)+E_{\rm p2}E_{\rm sig}\exp\Bigl(-2{\rm i}\omega_l\frac{z}{c}\Bigr).
\end{eqnarray} 
Since it is difficult to treat such fast oscillating phase terms in our numerical simulations, we introduce further approximations. 
In this experimental setup, the ac Stark shift term in equation (\ref{eq:Bloch3}) is small ($\max(|\Omega_{gg}-\Omega_{ee}|)=\mathcal{O}(1)$ MHz) compared to the laser linewidth and we ignore that term. 
Furthermore, $\rho_{ee}$ is at most $\mathcal{O}(10^{-5})$ and equation~(\ref{eq:Bloch3}) is approximated to
\begin{equation}
\frac{\partial\rho_{ge}}{\partial t} = -({\rm i}\delta+\gamma_{2})\rho_{ge} -{\rm i}\Omega_{ge}.\label{eq:Blochtmp}
\end{equation}
Finally, we separately consider fast oscillating terms in $\rho_{ge}$ and $\Omega_{eg}$ as
\begin{eqnarray}
\rho_{ge}=\rho_{ge}^{0}+\rho_{ge}^{+}\exp\Bigl(2{\rm i}\omega_l\frac{z}{c}\Bigr)+\rho_{ge}^{-}\exp\Bigl(-2{\rm i}\omega_l\frac{z}{c}\Bigr), \\
\Omega_{eg}=\Omega_{eg}^{0}+\Omega_{eg}^{+}\exp\Bigl(2{\rm i}\omega_l\frac{z}{c}\Bigr)+\Omega_{eg}^{-}\exp\Bigl(-2{\rm i}\omega_l\frac{z}{c}\Bigr), \\
\Omega_{eg}^{0}= \frac{\varepsilon_0\alpha_{eg}}{2\hbar}(E_{\rm p1}E_{\rm p2}+E_{\rm trig}E_{\rm sig}), \\
\Omega_{eg}^{+}=\frac{\varepsilon_0\alpha_{eg}}{2\hbar}E_{\rm p1}E_{\rm trig}, \ \ \  \Omega_{eg}^{-}=\frac{\varepsilon_0\alpha_{eg}}{2\hbar}E_{\rm p2}E_{\rm sig},
\end{eqnarray}
and each fast frequency component of equation~(\ref{eq:Blochtmp}) becomes
\begin{eqnarray}
 \frac{\partial\rho_{ge}^{0}}{\partial t} &=& -({\rm i}\delta+\gamma_{2})\rho_{ge}^{0} -{\rm i}\Omega_{eg}^{0*}, \label{eq:Blochs3} \\
  \frac{\partial\rho_{ge}^{+}}{\partial t} &=& -({\rm i}\delta+\gamma_{2}')\rho_{ge}^{+} -{\rm i}\Omega_{eg}^{-*}, \label{eq:Blochs4} \\
  \frac{\partial\rho_{ge}^{-}}{\partial t} &=& -({\rm i}\delta+\gamma_{2}')\rho_{ge}^{-} -{\rm i}\Omega_{eg}^{+*}. \label{eq:Blochs5} 
\end{eqnarray}
Each coherence component $\rho_{ge}^{0}$, $\rho_{ge}^{+}$, and $\rho_{ge}^{-}$ respectively represent coherence generated by 
the pump1 + pump2 and the trigger + signal fields, the pump2 + signal fields, and the pump1 + trigger fields.
The development of the population (equations (\ref{eq:Bloch1}, \ref{eq:Bloch2})) still include 
fast oscillating terms, but as shown in the next subsection the signal intensity does not depend on the population by adapting the approximation $\rho_{ee}\ll\rho_{gg}\simeq1$.   

In the simulations used in our previous experiments~\cite{bib:TPE3}, values of the $\gamma_2$ were calculated from Raman linewidths (equation (\ref{eq:Ramanwidth})) of an external experiment~\cite{bib:linewidth} as approximation. In this simulation we use the same value for the $\gamma_2'$ in equation (\ref{eq:Blochs4}) and (\ref{eq:Blochs5}). 
However, as described in section~\ref{sec:discussion1}, the Doppler broadening effect is negligible in the case of the excitation by the counter-propagating lasers. 
Thus, the value of the $\gamma_2$ in equation (\ref{eq:Blochs3}) is calculated from only the pressure broadening term of the Raman linewidth.

\subsection{Maxwell equations} 
We also consider developments of the laser fields and the macroscopic polarization of the p-H$_2$ molecules. 
Maxwell's equations in a homogeneous dielectric gas medium without sources and magnetization are given by
\begin{eqnarray}
\nabla\cdot\tilde{\bm{D}}=0, \ \ \  &&\nabla\cdot\tilde{\bm{B}}=0, \nonumber \\
\nabla\times\tilde{\bm{E}}=-\frac{\partial}{\partial t}\tilde{\bm{B}}, \ \ \ 
&&\nabla\times\tilde{\bm{B}}=\mu_0 \frac{\partial}{\partial t}\tilde{\bm{D}}, \nonumber \\
\tilde{\bm{D}}=\varepsilon_0\tilde{\bm{E}}+N_t\tilde{\bm{P}}, \ \ \ &&\nabla\cdot\tilde{\bm{E}}=0, \nonumber
\end{eqnarray}
where $N_t$ denotes the number density of the p-H$_2$ gas and $\tilde{\bm{E}}$ and $\tilde{\bm{P}}$ represent the laser fields and the macroscopic polarization of the p-H$_2$ molecules, respectively. 
From these equations, developments of the laser fields propagating in the $z$ direction are given by
\begin{equation}
\frac{\partial^2\tilde{\bm{E}}}{\partial t^2}-c^2\frac{\partial^2\tilde{\bm{E}}}{\partial z^2}=-\frac{N_t}{\varepsilon_0}\frac{\partial^2\tilde{\bm{P}}}{\partial t^2}. \label{eq:Maxwell0}
\end{equation}
In the current experimental setup, the $\tilde{\bm{E}}$ and the $\tilde{\bm{P}}$ are written by 
\begin{eqnarray}
\tilde{\bm{E}}=\tilde{\bm{E}}_{\rm p1}+\tilde{\bm{E}}_{\rm p2}+\tilde{\bm{E}}_{\rm trig}+\tilde{\bm{E}}_{\rm sig}, \\
-\tilde{\bm{P}}=-\bra{\psi}\bm{d}\ket{\psi} \label{eq:polarization} \\ 
=(c_{j-}^{*}c_g e^{{\rm i} \omega_{jg}t}d_{jg}+c_{g}^{*}c_{j+} e^{{\rm i} \omega_{gj}t}d_{gj} +c_{e}^{*}c_{j+} e^{{\rm i} \omega_{e'j}t}d_{ej}+c_{j-}^{*}c_e e^{{\rm i} \omega_{je'}t}d_{je})\hat{\bm{\epsilon}}_{R} + ({\rm c.c.}). \nonumber 
\end{eqnarray}
By using equations (\ref{eq:cjplus}) and (\ref{eq:cjminus}) and ignoring all the frequency components other than $\exp(\pm{\rm i}\omega_l t)$, 
equation (\ref{eq:polarization}) becomes
\begin{eqnarray}
\frac{2}{\varepsilon_0}\tilde{\bm{P}}&=&\Bigl[((\alpha_{gg}\rho_{gg}+\alpha_{ee}\rho_{ee})E_{\rm p1}+2\alpha_{eg}\rho_{ge}^{*}E_{\rm p2}^{*})\exp\left(-{\rm i}\omega_l\left(t-\frac{z}{c}\right)\right) \nonumber \\
&+& ((\alpha_{gg}\rho_{gg}+\alpha_{ee}\rho_{ee})E_{\rm sig}+2\alpha_{eg}\rho_{ge}^{*}E_{\rm trig}^{*})\exp\left(-{\rm i}\omega_l\left(t+\frac{z}{c}\right)\right) \nonumber \\
&+& ((\alpha_{gg}\rho_{gg}+\alpha_{ee}\rho_{ee})E_{\rm p2}^{*}+2\alpha_{eg}^{*}\rho_{ge}E_{\rm p1})\exp\left({\rm i}\omega_l\left(t+\frac{z}{c}\right)\right) \nonumber \\
&+& ((\alpha_{gg}\rho_{gg}+\alpha_{ee}\rho_{ee})E_{\rm trig}^{*}+2\alpha_{eg}^{*}\rho_{ge}E_{\rm sig})\exp\left({\rm i}\omega_l\left(t-\frac{z}{c}\right)\right)\Bigr]\bm{\hat{\epsilon}}_{R} \nonumber \\
&+&({\rm c.c.}).
\end{eqnarray}
By using following approximations
\begin{eqnarray}
\alpha_{gg}\rho_{gg}+\alpha_{ee}\rho_{ee}\simeq\alpha_{gg}, \\
\left(\frac{\partial}{\partial t}\pm{\rm i} \omega_l \right) (\rm{slowly\ varying\ term}) \simeq \pm{\rm i} \omega_l  (\rm{slowly\ varying\ term}) , \\
\left(c\frac{\partial}{\partial z}\pm{\rm i} \omega_l \right) (\rm{slowly\ varying\ term}) \simeq \pm{\rm i} \omega_l (\rm{slowly\ varying\ term}),
\end{eqnarray}
we obtain the developments of the envelopes of the electric fields:
\begin{eqnarray}
\left(\frac{\partial}{\partial t}-c\frac{\partial}{\partial z}\right) E_{\rm p1} &=& \frac{{\rm i}\omega_lN_t}{2}\left(\alpha_{gg}E_{\rm p1}+2\alpha_{eg}(\rho_{ge}^{0*}E_{\rm p2}^{*}+\rho_{ge}^{-*}E_{\rm trig}^{*})\right), 
\label{eq:Maxwells1} \\
\left(\frac{\partial}{\partial t}+c\frac{\partial}{\partial z}\right) E_{\rm p2} &=& \frac{{\rm i}\omega_lN_t}{2}\left(\alpha_{gg}E_{\rm p2}+2\alpha_{eg}(\rho_{ge}^{0*}E_{\rm p1}^{*}+\rho_{ge}^{+*}E_{\rm sig}^{*})\right), 
\label{eq:Maxwells2} \\
\left(\frac{\partial}{\partial t}-c\frac{\partial}{\partial z}\right) E_{\rm trig} &=& \frac{{\rm i}\omega_lN_t}{2}\left(\alpha_{gg}E_{\rm trig}+2\alpha_{eg}(\rho_{ge}^{0*}E_{\rm sig}^{*}+\rho_{ge}^{-*}E_{\rm p1}^{*})\right), 
\label{eq:Maxwells3} \\
\left(\frac{\partial}{\partial t}+c\frac{\partial}{\partial z}\right) E_{\rm sig} &=& \frac{{\rm i}\omega_lN_t}{2}\left(\alpha_{gg}E_{\rm sig}+2\alpha_{eg}(\rho_{ge}^{0*}E_{\rm trig}^{*}+\rho_{ge}^{+*}E_{\rm p2}^{*})\right). 
\label{eq:Maxwells4} 
\end{eqnarray}
As evident from these equations, signal photons are generated by the trigger field with the $\rho_{ge}^0$ 
and the pump2 field with the $\rho_{ge}^+$. 
Since the $\rho_{ge}^+$ is developed by the pump2 + signal fields and $|\rho_{ge}^+|$ is small, 
most part of the signal light is generated by the trigger field with the coherence generated by the pump beams.

\end{document}